\documentclass[a4paper]{article}
\usepackage{tabularx,booktabs,siunitx}
\usepackage{graphicx,subfig}
\usepackage{hyperref}
\usepackage{cleveref}
\usepackage{psfrag,authblk}

\begin{document}
	%opening
	\title{Interfacial free energy of the structure II cyclopentane hydrate in cyclopentane liquid from the capillary wave fluctuation method}
	\author{Bj{\o}rn Steen S{\ae}thre}
	\affil{Dept. of Physics \& Technology - University of Bergen\\P. O. Box 7803\\N-5020 Bergen, Norway\\E-mail: Bjorn.Sathre@ift.uib.no}
 \maketitle

\begin{abstract}
The interfacial stiffness\cite{Du2007} of structure II cyclopentane hydrate in cyclopentane liquid has been measured by the Capillary Wave Fluctuation (CWF) method for an equilibrated system with temperature of $T$=\SI{280}{K} and pressure of $P\sim$\SI{20}{bar}, close to the upper (L$_w$-Hyd-Vap-L$_{HC}$) Quadruple-point, on the boundary of the hydrate stable region. The hydrate-II/oil interfacial stiffness results fall broadly in the range $\Gamma_{ho}$=(35--55)$\times$\si{10^{-3}J/m^2}, with an average over all simulations giving\\ $\Gamma_{ho}$=\SI{45(5)d-3}{J/m^2}.  Earlier work has determined the cubic harmonic expansion of the natural crystalline oscillations in structure II hydrate\cite{Sathre2014}. Using this work, the stiffness value translates into an orientation-averaged interface free energy of $\gamma_{ho}$=\SI{45(5)d-3}{J/m^2}. The latter is in excellent agreement with the experimentally based estimates of $\gamma_{ho}$=\SI{47(5)d-3}{J/m^2} from particle-particle adhesion force measurements~\cite{Aman2013}.
This provides further evidence for the feasibility and robustness of the simulation approach to measure interfacial free energies in molecular systems.
\end{abstract}

\section{Introduction}
The interfacial free energy of gas hydrate interfaces~\cite{Zerpa2011}, both toward solid substrates~\cite{Kvamme_Kuznetsova2014,Aman2013b,Kumar2013,Aspenes_Dieker_Hoiland2010}, and fluids~\cite{Aman2013,Aman_Sloan2012,Aman2011,Aman2010}, are of crucial concern to evaluate both the likely site of hydrate initial growth~\cite{Kashchiev2002,Kashchiev2003} in process streams, and as input to meso-scale growth models~\cite{Tegze2006,Pusztai2008}. This parameter is also of crucial importance in understanding the detailed mechanism of kinetic hydrate inhibitors and anti-agglomerants (AAs). It has the potential to elucidate if a surrounding oil phase is needed for AA action, or whether a coating of surface active components a few (tens of) molecular layers thick is sufficient to effectively screen the hydrate-hydrate adhesion forces. Recent experiments may suggest the latter~\cite{Sun_Firoozabadi2013}.
A general simulation scheme for determining the interfacial free energy of hydrate interfaces, could become part of model-based screening for potential new compounds for use in hydrate flow assurance. It could also be useful in risk assessment programs to find potential physical locations where hydrate growth is probable. Conversely, in hydrate/mineral interfaces it can be used to help evaluate stability properties of and kinetic barriers to hydrate extraction in model reservoirs. 

\section{Methodology}
The physics, the simulation procedure and the analysis algorithm have been described in our previous work~\cite{Sathre2014}, and only a brief summary will be given here.
Molecular dynamics simulations of elongated rectangular,slab-like hydrate/fluid topologies were conducted to detect and quantify the CWFs in the interface~\cite{Hoyt2010}.
The interfaces have a pseudo-1D geometry with the bi-tangential length significantly shorter than the tangential one. This is done to enhance the magnitude of the CWFs~\cite{Gelfand1990}. 
The molecular dynamics package GROMACS~\cite{Hess2008b} was applied, with the utilization of the OPLS$_{UA}$ forcefield~\cite{Jorgensen1984,Jorgensen1996} for the cyclopentane (CPN) molecules, and the compatible TIP4P/ice~\cite{Vega2005} rigid, non-polarizable model for the water molecules. 
The equations of motion were integrated by the Leap-frog half-step scheme. A Cartesian simulation box was used with periodic boundary conditions in all 3 dimensions. 
The TIP4P/ice model was augmented with the Particle Mesh Ewald(PME) scheme~\cite{Essmann1995} for long range electrostatics. The parameters used were a k-space cutoff of \SI{0.85}{nm} and 4th order B-spline interpolation. This is now the preferred approach also with rigid water models~\cite{Vega2011}, despite the fact that The TIP4P/ice model is parameterized for plain cut-off electrostatics.
A single cut-off of \SI{0.85}{nm} was used for Van der Waals interactions. The water model was kept rigid by the standard SETTLE~\cite{Miyamoto1992} algorithm, whereas the CPN-bonds were kept rigid during simulations by the P-LINCS~\cite{Hess2008} scheme, using a 4th order matrix series expansion and a single iteration. 
After an annealing scheme to minimize the hydrate-slab dipole moments~\cite{Sathre2012}, and a separate equilibration of the oil phase, the two phases were brought into contact.  
A further equilibration with the following settings were conducted on the merged hydrate/CPN(l) system:
\subsection{Equilibration settings}
A total of \SI{5}{ns} were run with a time step of \SI{2}{fs}, after which the energy and pressure stabilized.
Neighbor searching was performed every 5 steps.
The PME algorithm was used with an initial k-space grid of 320 x 15 x 512 cells.
Temperature and a semi-isotropic pressure control were done with the Berendsen scheme~\cite{Berendsen1984,Amadei1998}
with time constants, $t(T)$=\SI{0.5}{ps} and $t(P)$=\SI{2}{ps} respectively.
Due to the presence of the interface, the tangential and normal pressures were adjusted sequentially, first the normal and then the tangential pressure to approximate the procedure in~\cite{Frenkel2013}.
The compressibilities of the barostat were $\kappa_{\mathbf{t}}$=\SI{1.2d-5}{bar^{-1}} and $\kappa_{\mathbf{n}}$=\SI{8d-5}{bar^{-1}} for the directions tangential and normal to the interface respectively. 
\subsection{Simulation settings}
A total of \SI{3}{ns} were simulated with a time step of \SI{2}{fs}.
Neighbor searching was performed every 10 steps.
The PME algorithm was used with a k-space grid of 240 x 12 x 384 cells.
Temperature coupling was done with the V-rescale algorithm~\cite{Bussi2007}, enforcing the canonical ensemble with the same time-constant as the equilibration.

\subsection{Analysis}
The analysis was identical to our previous work~\cite{Sathre2014} and consisted of two stages:
\begin{enumerate}
\item Restricting attention to the interface region along the normal-to-interface coordinate. 
\item Applying a Gibbsian dividing surface criterion to the order parameter profile in the interface region, as delimited in 1.
\end{enumerate}

Stage 1 is achieved by fitting the order parameter profile to a suitable interface functional, typically a hyperbolic tangent, and using the location and width parameters of the functional to delimit a specific interval along the normal(-to-interface) coordinate\cite{Vink_Horbach2005}. This procedure is repeated for every trajectory snapshot so it is possible to correct for moving interfaces in the course of longer simulations.
This scheme helps eliminate the resolution limitation that would otherwise be incurred by a rectangular spatial mesh of the order parameter in the box.
The mass density was used as order parameter throughout.

\section{Results}
\begin{figure}[htb]
\subfloat[Initial configuration]{
\includegraphics[bb=0 0 90 144]{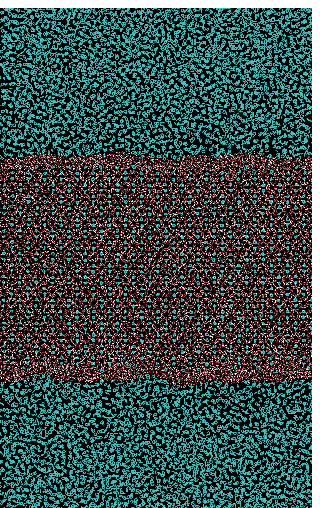}}
% hydII_110_3.pdf: 90x144 pixel, 72dpi, 3.17x5.08 cm, bb=0 0 90 144
\hspace{6pt}
\subfloat[After liquid bridge formation]{\includegraphics[bb=0 0 90 144]{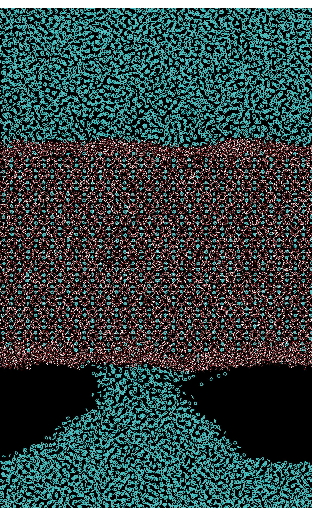}}
% hydII_110_lb.pdf: 90x144 pixel, 72dpi, 3.17x5.08 cm, bb=0 0 90 144
\caption{Snapshots of system configuration of simulation no. 4 viewed along the [110]-direction , H2O (red), Cyclopentane (blue) - symmetrized around the hydrate phase for clarity.}
\label{fig:setup}
\end{figure}

The simulation configurations are detailed in \cref{tab:simsetup}. The use of the rectangular slab topology creates two interfaces in the box, which we denote the upper and lower interface according to their normal(-to-interface) coordinate. A snapshot of one simulation in its initial stage is shown in \cref{fig:setup}a.
Energy statistics and thermodynamics in the equilibration and simulation stages are given in \cref{tab:thermodyn}. The results show that the average pressure is well adjusted in the equilibration stage. Starting from a particular equilibrated state, with no barostat applied, however, leads to large deviations from the set pressure, reflecting the inherent pressure fluctuations around the average.  This is particularly an issue with our extremely thin (pseudo-1D) slab topology, even in a multistage pressure relaxation. This discrepancy may indicate that our approach to approximate the scheme of pressure equilibration in~\cite{Frenkel2013} is not completely satisfactory.
The CFM spectra with best fit-lines for Simulation no. 3 are given in \cref{fig:sim280}.
The resulting fit diagnostics and estimated stiffnesses are shown in \cref{tab:fitdiag}.
The upper and lower interfaces do not show identical physical behavior, as also reflected in the results: The two simulations at $T$=\SI{280}{K} yield surface stiffnesses of $\Gamma_L$=\SI{52(8)d-3}{J/m^2} and $\Gamma_U$=\SI{42(8)d-3}{J/m^2}.     
 The lower interfaces tend in some cases to have liquid bridge formation of cyclopentane with its own vapor.
This bridge formation (\cref{fig:setup}b) occurs spontaneously within \SI{2}{ns} in most simulations. Only the upper interfaces remain fully immersed in cyclopentane throughout.
The results from the upper and lower interface are both in agreement within $1\sigma$ with the result $\gamma_{ho}$=\SI{47(5)d-3}{J/m^2} from particle-particle adhesion experiments~\cite{Aman2013}, when taking into consideration the small difference in the values of stiffness and interfacial free energy. This difference is no greater than \SI{0.3d-3}{J/m^2} as estimated from the fit to cubic harmonics in the hydrate-II/vacuum simulations\cite{Sathre2014}.
Due to the liquid bridge formation, the most reliable results are likely to be those from the upper interfaces in Sim. no. 3 and 4.

\section{Discussion/Conclusion}
\begin{figure}[htb]
\subfloat[Upper interface]{
%\begin{postscript}
\psfrag{X}{k[nm$^{-1}$]}
\psfrag{Y}{Power[nm$^2$]}
\includegraphics[bb=0 0 180 115]{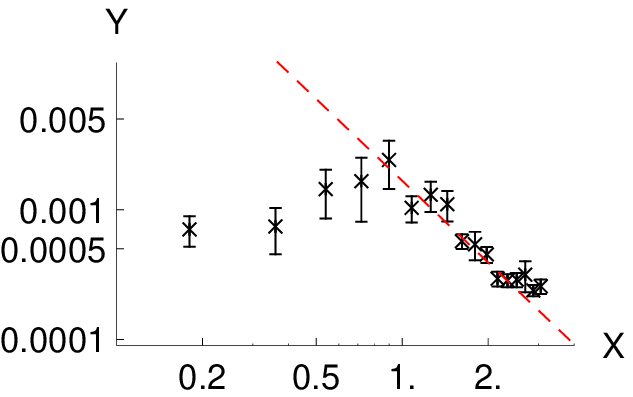}}\\
% spect_280K_20x1x10_tr2_a3_int2.eps: 0x0 pixel, 600dpi, 0.00x0.00 cm, bb=0 0 180 115
%\end{postscript}
\subfloat[Lower interface]{
%\begin{postscript}
\psfrag{X}{k[nm$^{-1}$]}
\psfrag{Y}{Power[nm$^2$]}
\includegraphics[bb=0 0 180 115]{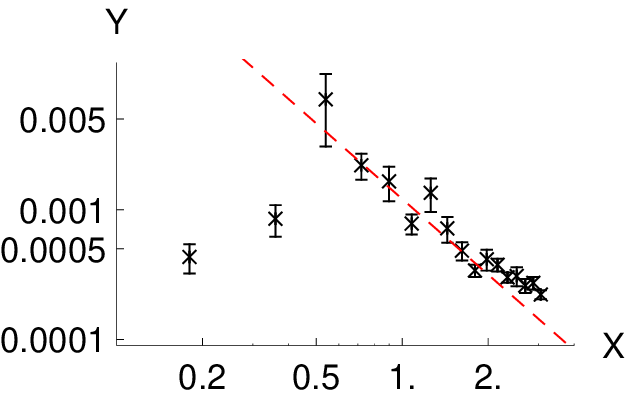}}
% spect_280K_20x1x10_tr2_a3_int1.eps: 0x0 pixel, 600dpi, 0.00x0.00 cm, bb=0 0 180 115	
%\end{postscript}}
\caption{Log-log plotted Capillary Wave spectra for Sim. no. 3, $T$=\SI{280}{K}, $\alpha$=3 with best-fit lines dashed in red}
\label{fig:sim280}
\end{figure}

The scatter in the results is appreciable and many of the spectra show significant noise contamination. As in our previous work on the same simulation scheme~\cite{Sathre2014} there are significant interface-interface dynamical correlations, even though the positive and negative cross-correlations largely average out over the whole interfacial strip, as seen in \cref{tab:intfluctcorr}. (The dispersion of the cross-correlation values are larger than their average). These correlations are indicative of interface-interface interference which plays a major role in attenuating the lower harmonics seen in the spectra \cref{fig:sim280}a,b. This can be avoided only by extending the system size in the normal(-to-interface) direction\cite{Sathre2014}.
The systematic differences in the free energy between the upper and lower interfaces have already been noted. The reason why cyclopentane liquid bridges form only at the lower interface in our simulations is unclear. A possible reason may be that energy is not intrinsically independent of normal coordinate in the box: In the trajectories, the lower interfaces essentially coincide with the simulation box lower boundary. This energy effect could then perhaps be due to artifacts of Ewald sum electrostatics. However, we do not have sufficient data to discount the effect as being an expression of spontaneous symmetry breaking from a saddle point in the potential, rather than a fully reproducible outcome from the initial conditions. Due to the constant volume and particle number it is observed that a formed liquid bridge will likely preclude the formation of another on the opposite hydrate/oil interface.
 
Further visual inspection of the trajectories indicated some surface melting of the hydrate as the capillary waves rippled across the interfaces. 
It was generally straightforward to judge by a combination of visual inspection and potential energy change which results were reliable and which were too contaminated to be trusted.

The melting line of the CPN-hydrate/liquid simulation system is almost identical to the near-vertical L$_W$-Hyd-L$_{HC}$ experimental 3-phase line in the P,T phase diagram, starting from the second quadruple point at $(T,P)$=(\SI{281}{K},\SI{1}{bar})\cite{Aman2011}. We subtract $\Delta T$=\SI{1}{K}, due to the fact that the TIP4P/ice water model has ice point $T$=\SI{272.1}{K}. 
It is clear that in temperatures lower than $T$=\SI{280}{K} the capillary wave magnitudes remain small, the sufficiently large ones occurring only close to the phase transition temperature. The exception to this is the lower interface of Sim. no. 1, where a large liquid bridge is formed, and therefore the effective phase transition temperature is close to $T$=\SI{277}{K} - the melting point of CPN-hydrate in vacuum/vapor. This gives a free energy somewhere between that of hydrate/vacuum and hydrate/oil on this interface.

We can deduce from the above that, in the absence of extensive liquid bridge formation, the capillary wave method can function with a temperature fixed at or just above the hydrate/fluid melting line, provided the system remains stable (in potential energy) during data collection, but it does not function well with a temperature set significantly below the melting point of the system at the imposed pressure. 
Depending on the temperature increase over the phase transition point, it can be questioned, whether we in such a semi- or pseudo-stable system really measure the interfacial free energy of the solid/liquid system as opposed to a liquid/liquid or adsorbed-phase/liquid system. 
It is a validation of the robustness of the method that despite all these difficulties, even in the presence of some liquid bridging, the average value obtained from all simulations close to the melting line at $T$=\SI{280}{K} show such a significant overlap with the best experimental results on this system~\cite{Aman2013}.

\clearpage
\begin{table}
\begin{tabular}{llrrrr}
\toprule
	&	   &		&\multicolumn{3}{c}{Size (after equilibration)}\\	
\midrule							   
Sim. no.&HydII UC&N(CPN) 	&Lx[nm]		&Ly[nm]	&Lz[nm]\\
\midrule
1	&20x1x10   &16900	&35.7972	&1.7945	&57.7884\\ 
2	&20x1x10   &16920	&35.7972	&1.7945	&57.7884\\ 
3	&20x1x10   &16920	&34.9242	&1.7443	&58.2437\\   
4	&20x1x10   &16920	&34.9156	&1.7477	&56.6836\\	
\bottomrule
\end{tabular}
\caption{Simulation setup}
\label{tab:simsetup}
\end{table}

\begin{table}
\begin{tabular}{lrSSSS}
\toprule
Sim. no.&t$_{traj}$[ps]	&T[K]		&P[bar]		&Mu[{D}]	&{E$_{pot}$}[kJ/mol]\\
\midrule
1-2-EQ	&4000		&274.96(2)	&19.5(8)	&790(30)&-33.489(4)\\ 
3-EQ	&5000		&279.945(4)	&19.3(3)	&650(40)&-33.18(5)\\  
4-EQ	&5000		&279.95(1)	&20.0(2)	&670(30)&-33.4138(4)\\
1	&3000		&277.021(9)	&-329(2)	&855(2)	&-33.5459(2)\\
2	&3000		&279.01(1)	&-316.9(3)	&770(14)&-33.4774(2)\\
3	&1740		&279.77(1)	&-2.8(9)	&638(3)	&-33.30(6)\\  
4	&3000		&279.93(5)	&-99.7(4)	&658(3)	&-33.97(1)\\ 
\bottomrule
\end{tabular}
\caption{Equilibration and Simulation thermodynamics}
\label{tab:thermodyn}
\end{table}

\begin{table}
\begin{tabular}{lrrrSSS}
\toprule 
Sim. no. &T[\si{K}]&Int. &\# Pts. &Slop{e}	&{Adj. R$^2$}&$\Gamma$[\si{10^{-3}J m^{-2}}]\\
\midrule		
1	&277    &L	&14	&-2.13(21)	&0.883	&57(10)\\
       	&  	&U	&14	&-1.79(24)	&0.802	&127(23)\\
2	&279	&L	&8	&-1.09(66)	&0.201	&127(25)\\
	&	&U	&14	&-1.79(24)	&0.802	&217(39)\\
3	&280	&L	&9	&-1.94(26)	&0.876	&53(7)\\
	&	&U	&9	&-2.09(23)	&0.914	&38(6)\\
4	&280	&L	&12	&-2.12(26)	&0.856	&50(8)\\
	&	&U	&9	&-2.12(40)	&0.771	&46(9)\\
\bottomrule
\end{tabular}
\caption{Fit diagnostics and resulting interfacial stiffnesses for the simulations in this work, at $\alpha$=3}
\label{tab:fitdiag}
\end{table}

\begin{table}
\begin{tabular}{lS}
\toprule
Sim. no. & {Avg.Corr.}\\
\midrule
1 & 0.0(2)\\
2 &-0.1(2)\\
3 & 0.0(4)\\
4 & 0.0(3)\\
\bottomrule
\end{tabular}
\caption{Interface-interface position cross-correlations - averaged over all in-interface positions}
\label{tab:intfluctcorr}
\end{table}

\end{document}